\documentclass{ws-procs9x6}

\begin{document}

\title{The X-ray properties of the PG QSO sample 
observed with XMM-Newton}

\author{E. JIM\'ENEZ, E. PICONCELLI, M. GUAINAZZI and
N. SCHARTEL}

%\footnote{\uppercase{XMM-Newton Science Operations Center, RSSD of ESA, VILSPA, Madrid, Spain}}

\address{XMM-Newton Science Operations Center, RSSD of ESA, \\ 
Apd 50727, 28080-MADRID SPAIN \\ 
}


\maketitle

\abstracts{
We  present  preliminary results   of a   systematic analysis  of  the
XMM-Newton spectra of nearby optically  bright QSOs. The objects  have
been selected  from the Bright Quasar Survey  sample. The goal of this
project is to characterise the X-ray  spectral properties of optically
selected QSOs in  the  0.3--12 keV  energy  band.  In most cases,  two
component continua are    sufficient  to model  reasonably  well   the
observed spectra.  All  but  one  detected  Fe K$\alpha$  line have  
narrow profiles.
}

\section{Preliminary results of the analysis of the XMM-Newton PG QSO sample}

The sample  consists of 30 QSOs  observed by  XMM-Newton, and selected
from the Bright Quasar Survey,  (M$_{\rm B}<$-23), representing  ~25\%
of the total catalogue.  The redshift of the  objects ranges from 0.04
to  1.72 and    the Galactic  equivalent   column density   is  below
$5.7\times10^{20}$~cm$^{-2}$.  The majority of the sources, (27 out of
30), are classified as Radio Quiet QSOs. The X-ray luminosity covers a
range of $5\times10^{43}<L_{2-10keV}< 5\times10^{45}$ erg/s.

\begin{figure}[ht]
%\epsfxsize=10cm   %width of figure - will enlarge/reduce the figures
%\epsfbox{fig3.eps}
%\figurebox{2cm}{3cm}{} %to have a box alone 
%\centerline{%\epsfysize=8cm\epsfbox{g_lum.eps}
%\epsfysize=8cm\epsfbox{bw.ps}}
\includegraphics[width=8cm, angle=90]{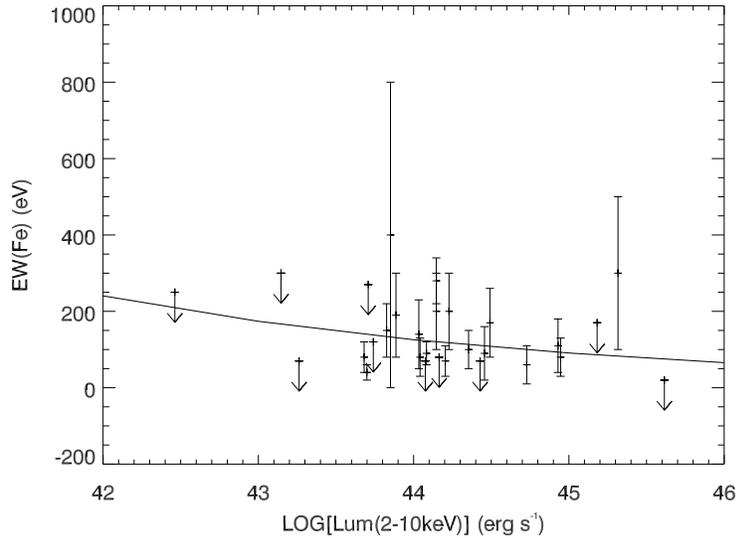}
\caption{X-ray Baldwin effect: equivalent width of the narrow Fe line decreases
with hard X-ray luminosity, EW    $\propto  L^{-0.15\pm0.08}$. \label{figs}}
\end{figure}

The continua of all QSOs are satisfactorily fitted by a power law plus
a  soft excess  component.   The mean value   of the hard X-ray photon
index  of the  power law  is  $\Gamma_h=2.0\pm0.2$, fully in agreement
with ASCA  (George et al.    2000, {\it ApJ}    {\bf 531}, 52),  GINGA
(Lawson  \& Turner  1997,  {\it  MNRAS} {\bf   288}, 920) and   EXOSAT
(Comastri et al.   1992, {\it ApJ}   {\bf 384}, 52)  previous results.
$\Gamma_h$ is independent  of 2--10 keV  luminosity; however,  a large
scattering around the mean value  is clearly detected.  Four different
spectral  parameterizations have been tested to   account for the soft
excess emission,  i.e.  a {\it power  law}, a  {\it black body},1 {\it
multi-blackbody disk} and a {\it thermal  Bremsstrahlung} model.  In 8
out of 30 QSOs two blackbodies are necessary.  We also checked for the
presence of  other spectral features both in  the hard and soft bands.
The presence  of soft excess  is ubiquitous and  it was fitted with: a
{\it black body} model in 12 objects with $<$kT$>$= 0.15$\pm$0.04~keV;
a combination   of   2  {\it   black   bodies}   in   8  objects  with
$<$kT$_1$$>$=0.11$\pm$0.08~keV  and $<$kT$_2$$>$=0.28$\pm$0.03~keV;  a
{\it       bremsstrahlung}     model      in     7     objects    with
$<$kT$>$=0.43$\pm$0.02~keV; a  {\it power law}  component in 2 objects
with $<$$\Gamma$$>$=2.5$\pm$0.8.   A  {\it multi-temperature blackbody
disk} emission does not fit the soft excess in any QSO.  In 5 objects,
the  presence of a  complex absorption pattern in  the soft X-ray band
requires a multi-component ionized absorber.  Furthermore, single soft
X-ray  edges at $\sim$0.74  keV likely  due to OVII  or  UTA have been
observed  in  14 out of   30  QSOs.  16  QSOs  show  evidence of a  Fe
K$_\alpha$ emission line.  All but one detected lines present a narrow
profile.  The    mean  line energy  is  E=6.4$\pm$0.2  keV.     The Fe
K$_\alpha$  shell   is dominated by  emission from   cold material (Fe
I-XVI).  No significant trend in the line  energy as a function of the
2--10  keV luminosity has been   found.   An X-ray Baldwin effect   is
observed: the equivalent    width EW of the  Fe    line decreases with
increasing     luminosity,   EW    $\propto  L^{-0.15\pm0.08}$,    (see
Figure~\ref{figs}).    The value of  the   slope is  in agreement  with
Iwasawa \& Taniguchi (1993,  {\it ApJ} {\bf  413}, 15) and Page et al.
(2003, {\it astro-ph/0309394}).

%\section{Mrk~304}

%Mrk 304 shows a complex spectrum, dominated by heavy (N$_{\rm{H}}\sim$
%10$^{23}$~cm$^{2}$)  ionized obscuration.   A two-phase warm  absorber
%provides  a good description for the  absorbing plasma, (Piconcelli et
%al. 2004 {\it MNRAS} accepted).  Our results appear to be in agreement
%with a  multi-zone structure for  the ionized  gas.   This scenario is
%supported by X-ray high  resolution spectroscopy of bright AGNs, where
%the  complex structure   of   the circumnuclear  ionized  material  is
%resolved  into a  multi-phase plasma with  large spread  in $\xi$  and
%systematic blue shifts  (Krolik 2002, astro-ph/0204418).  Two emission
%features have been significantly detected at  ~0.5 keV and ~6 keV. The
%former is likely due to a He-like oxygen triplet at 0.57 keV, probably
%originating in the same warm medium responsible  for the absorption in
%the soft band.   The latter is due to  fluorescent emission of K-shell
%iron.   The  combination  of  the   energy  centroid  ($\sim$6.4 keV)   and
%$\sigma_{K_\alpha}  <$ 0.18 keV  suggests an origin from cold (FeI-XV)
%matter  distant from the  inner  disk region.   Finally, we report the
%presence of a soft  excess component (with normalization $\sim$4\%  of
%the primary emission at 1 keV).  This emission  could be due to either
%partial covering or scattered emission   from the ionized   outflowing
%plasma.

\end{document}